\documentclass[twocolumn,showpacs,amsmath,amssymb]{revtex4}
\usepackage[T1]{fontenc}
\usepackage[latin1]{inputenc}
\usepackage{epsfig}
\usepackage[bf]{caption}
\usepackage{graphicx}
\usepackage{dcolumn}
\usepackage{bm}

\newcommand{\be}{\begin{equation}}
\newcommand{\ee}{\end{equation}}
\newcommand{\ba}{\begin{eqnarray}}
\newcommand{\ea}{\end{eqnarray}}

\begin{document}
\title{Influence of Random Internal Fields on the Tunneling of OH$^-$ Defects in NaCl Crystals}
\author{M. Thesen}

\author{R. K\"uhn}
\affiliation{Institut f\"ur Theoretische Physik, Universit\"at Heidelberg, D-69120 Heidelberg, Germany}

\author{S. Ludwig}
\author{C. Enss}
\affiliation{Kirchhoff-Institut f\"ur Physik, Universit\"at Heidelberg D-69120 Heidelberg, Germany}

\date{\today}


\begin{abstract}
Alkali halide crystals doped with certain impurity ions show a low temperature behaviour, which differs significantly from that of pure crystals. The origin of these characteristic differences are tunneling centers formed by atomic or molecular impurity ions. We have investigated the dielectric susceptibility of hydroxyl ions in NaCl crystals at very low concentrations (below 30 ppm), where interactions are believed to be negligible. We find that the temperature dependence of the susceptibility is noticeably different from what one would expect for isolated defects in a symmetric environment. We propose that the origin of these deviations are random internal strains arising from imperfections of the host crystal. We will present the experimental data and a theoretical model which allows a quantitative understanding on a microscopic basis. 
\end{abstract}

\maketitle


For almost 30 years phenomenological tunneling models have provided a powerful conceptual framework for the understanding of both amorphous solids \cite{Phil}\cite{An} and point defects in crystals \cite{NaPo}. It has been a widely accepted belief that in the limit of large impurity concentration the disordered crystals show glasslike behaviour. One of the main interests in studying crystals with varying impurity concentrations has been to observe the onset of glasslike low-temperature universality  \cite{WaPo}.

In this work we want to step back and investigate the dynamics of supposedly non interacting hydroxyl defects at low concentrations in ${\rm NaCl}$ crystals. A thorough description of the properties of this type of material could then serve as a starting point for the modelling of highly doped crystals with stronger disorder where the interaction has to be included. 


We have investigated the dielectric susceptibility $\chi(\omega)$ of NaCl:OH$^-$ at several low impurity concentrations and at frequencies ranging from 100 Hz to about 1 GHz and temperatures from $0.1$ to $10$ K. A single ${\rm OH^-}$ - ion replacing a ${\rm Cl^-}$ can be described microscopically by a simple six -- state model \cite{Shore}. The results show clear differences from the naive expectation based on this microscopic picture (cf. Fig. 1). Especially in the crossover region at temperatures comparable to the tunnel splitting the dielectric susceptibility $\chi'$ varies more smoothly with temperature than expected. This holds for all frequencies and concentrations observed. 

\begin{figure}
\begin{minipage}[t]{0.5\textwidth} 
\includegraphics[width=\textwidth]{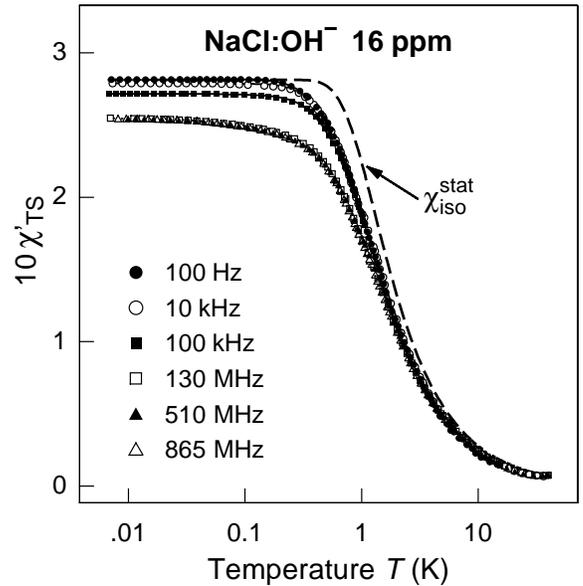}
\caption{Real part $\chi'$ of the dielectric susceptibility of ${\rm OH^-}$ doped ${\rm NaCl}$ at several frequencies. The dashed line represents the theoretical expectation for isolated defects.}
\end{minipage}
\end{figure}

In addition, the damping, $\chi''(\omega)$, is surprisingly high and changes within a frequency range where we would have expected the behaviour to be well described by the static limit $\omega \to 0$, in which no dissipation should occur. This phenomenon has been discussed in terms of interacting defects \cite{Wu}\cite{PoAn}.   In this work we want to restrict ourselves to isolated systems and focus on the course of the susceptibility curve.


Classically, the egg-shaped ${\rm OH^-}$ -ion in a NaCl crystal possesses six equilibrium positions between which it can move via tunneling processes. This partially lifts the sixfold degeneracy of the groundstate. The defect is thus modelled by a six-state system consisting of pocket states which are supposed to be highly localized and to be eigenstates of the position operator $\vec X$ with eigenvalues corresponding to the positions of the six minima of the potential energy. We  assign a tunneling matrix element $-a$ to the $90^0$ - tunneling and a matrix element $-b$ to the $180^0$ - processes. This leads to three energy levels $E_{\rm g}=-4a-b$, $E_{\rm a}=b$ and $E_{\rm d}=2a-b$ with degeneracies  1,3, and 2 respectively. The triplet consists of states with negative parity, whereas the other states have positive parity, such that no dielectric transitions are allowed between the levels $E_{\rm g}$ and $E_{\rm d}$.  
 
The total Hamiltonian of a single defect in an electric field $\vec E$ and an elastic strain -- field reads
\[ H=H_{\rm tunnel}+q \vec X \cdot \vec E + \gamma \sum_i \epsilon_{ii}X_i^2\]
where the off- diagonal elements $\epsilon_{ij}\quad i\neq j $ of the strain tensor cannot couple to the defect because of the operators $X_i$ and $X_j$ projecting onto different and orthogonal states. Here, $q$ is the effective charge of the defect and $\gamma$ describes the coupling to strain fields and can be extracted from measurements of the sound velocity and -attenuation. For a detailed description the reader is referred to \cite{Shore}

In the presence of elastic strain fields, the selection rules mentioned above remain unchanged  whereas electric fields will mix the parity eigenstates, and dielectric transitions between all the states enter the game. As a consequence, static electric fields lead to an augmentation of the dielectric susceptibility at temperatures around $2a/k_{\rm B}$, whereas the main effect of elastic fields is the spreading of the spectrum and a consequent flattening of the susceptibility vs. temperature curve.

From the above mentioned deviations from the naive expectation, we conjecture that the OH$^-$ defects in alkali halide crystals are subject to random internal strain fields in the host crystal, whereas random electric fields do not seem to be of great importance. 
Not knowing the origin of these strain fields, we suppose them to be gaussian distributed with zero mean and variance $\sigma$ and calculate the observed $\chi$ by averaging over this distribution.

We find a very nice agreement with the experimental data using for the sample containing 16 ppm OH$^-$ the parameters $\sigma = 1.57$ $k_{\rm B}$K, $2a=1.29$ $k_{\rm B}$K and $b=0$ where the latter is in fair agreement with preceding experiments ( See, e.g. \cite{Suto}). With these parameters our model fits the real part of the static susceptibility very well (cf. Fig. 2).

\begin{figure}
\begin{minipage}[t]{0.5\textwidth}
\includegraphics*[width=\textwidth]{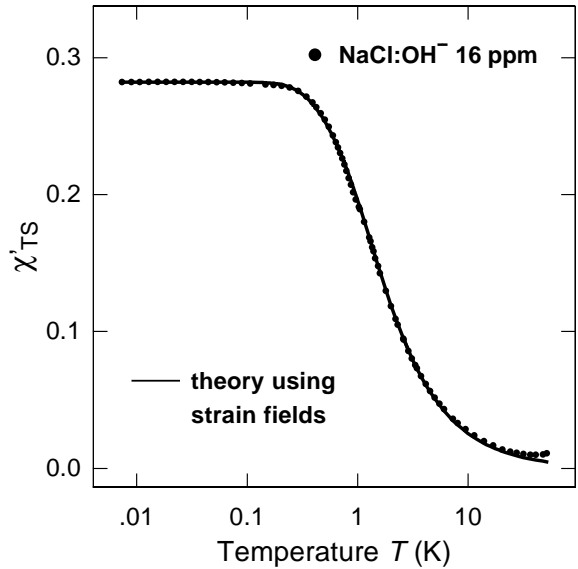}
\caption{Dielectric susceptibility $\chi'$ of NaCl depend with 16ppm OH$^-$ at 10 kHz and fit (solid line) using random internal strain field distributed with variance $\sigma = 1.57 k_{\rm B}$K. The $90^0$ tunneling parameter $a$ has been found to be $a=1.29k_{\rm B}$K.}
\end{minipage}
\end{figure}


The frequency dependence of $\chi'$ and the behaviour of the damping $\chi''$ indicates the presence of additional relaxation channels. 

As relaxation through the phonon bath would be too weak to account for the observations (cf. Fig. 3), we conjecture that relaxation is mainly due to direct interactions between the defects, even at concentrations as low as 16 ppm.

\begin{figure}
\begin{minipage}[t]{0.5\textwidth}
\includegraphics[width=\textwidth]{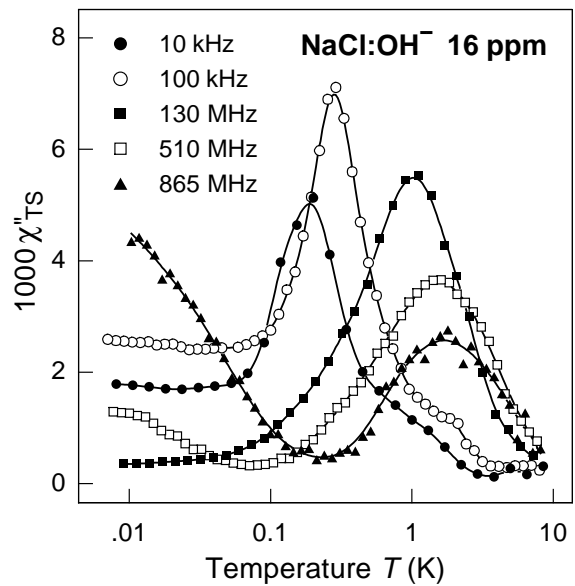}
\caption{Imaginary part of the dielectric susceptibility $\chi''$ of ${\rm OH^-}$ doped ${\rm NaCl}$ at several frequencies.}
\end{minipage}
\end{figure}

The most probable candidates for the possible sources of the strain fields are dislocation lines within the host crystal. With $10^8$ ${\rm cm}^{-2}$ as a reasonable value for their concentration each defect would have an dislocation line at an average distance of about 10 \AA. If we plug in appropriate values for the coupling- and the lattice constant we find an average interaction energy of 5 $k_{\rm B}$K which agrees with our previous estimates. 

In measurements of the temperature dependence of the dielectric constant of ${\rm OH}^-$  doped alkali halide crystals we have observed deviations from the behaviour one would have expected for isolated point defects in a symmetric environment. Investigation of the influence of random strain- and electric field led us to the conclusion that the experiment proved the presence of random strain fields in the host crystal. The order of magnitude of the strain fields needed to explain the experiment can easily be covered by a reasonable concentration of displacement lines.

\end{document}